\def\e{{\rm e}}
\def\del{\partial}
\def\half{{1\over2}}
\def\cosb{\cos\beta}
\def\sinb{\sin\beta}
\def\z{\zeta}
\def\abs#1{{\left|{#1}\right|}}

\newcommand{\PRD}[3]{Phys. Rev. {\bf D{#1}} (19{#2}) {#3}}

\newcommand{\NPB}[3]{Nucl. Phys. {\bf B{#1}} (19{#2}) {#3}}
\newcommand{\PLB}[3]{Phys. Lett. {\bf B{#1}} (19{#2}) {#3}}
\newcommand{\PTP}[3]{Prog. Theor. Phys. {\bf {#1}} (19{#2}) {#3}}

\documentstyle[12pt,epsf]{article}
\makeatletter

\@addtoreset{equation}{section}
\makeatother
\begin{document}
\begin{titlepage}
\begin{flushright}
SAGA--HE--89\\
YITP/K--1115\\
July 29,~1995
\end{flushright}
\vspace{24pt}
\centerline{\Large {\bf CP-Violating Profile of the Electroweak Bubble Wall}}
\vspace{24pt}
\begin{center}
{\bf Koichi~Funakubo$^{a,}$\footnote{e-mail: funakubo@cc.saga-u.ac.jp},
 Akira~Kakuto$^{b,}$\footnote{e-mail: kakuto@fuk.kindai.ac.jp},
 Shoichiro~Otsuki$^{b,}$\footnote{e-mail: otks1scp@mbox.nc.kyushu-u.ac.jp}\\
 Kazunori~Takenaga$^{c,}$\footnote{e-mail: takenaga@yukawa.kyoto-u.ac.jp}
 and Fumihiko~Toyoda$^{b,}$\footnote{e-mail: ftoyoda@fuk.kindai.ac.jp}}
\end{center}
\vskip 0.8 cm
\begin{center}
{\it $^{a)}$Department of Physics, Saga University,
Saga 840 Japan}
\vskip 0.2 cm
{\it $^{b)}$Department of Liberal Arts, Kinki University in Kyushu,
Iizuka 820 Japan}
\vskip 0.2 cm
{\it $^{c)}$Yukawa Institute for Theoretical Physics,
Kyoto University, Kyoto 606--01 Japan}
\end{center}
\vskip 1.0 cm
\centerline{\bf Abstract}
\vskip 0.2 cm
\baselineskip=15pt
In any scenario of the electroweak baryogenesis, the profile of the
CP violating bubble wall, created at the first-order phase transition,
plays an essential role. We attempt to determine it
by solving the equations of motion for the scalars in the two-Higgs-doublet
model at the transition temperature.
According to the parameters in the potential, we found three solutions.
Two of them smoothly connect the CP-violating broken phase and 
the symmetric phase, while the other connects CP-conserving vacua but
violates CP in the intermediate region.
We also estimate the chiral charge flux, which will be turned into the
baryon density in the symmetric phase by the sphaleron process.
\end{titlepage}
\baselineskip=18pt
\setcounter{page}{2}
\setcounter{footnote}{0}
%
%
%\section{Introduction}
\section{\protect{\large\bf Introduction}}
Since the fascinating proposal of the electroweak baryogenesis\cite{reviewEB},
its various aspects have been investigated by many authors.
For the baryogenesis to occur, the following conditions must be met:
(1) The electroweak phase transition (EWPT) must be first order to realize
a state out of equilibrium. (2) There had to be CP violation in the era of the
EWPT. Besides these, to keep the generated baryons, (3) the sphaleron processes
must decouple just after the EWPT.
These impose some restrictions on the models of the electroweak theory.
The conditions (1) and (3) give an upper bound on the lightest neutral Higgs
particle, which is inconsistent with the present lower bound of the Higgs
scalar in the minimal standard model. Further to have efficient CP violation,
an extension of the Higgs sector would be needed.
On this ground, the two-Higgs-doublet model, including the minimal
supersymmetric standard model (MSSM), has been studied to estimate 
the generated baryon number.\par
It is essential in the scenarios of the baryogenesis,
spontaneous or charge transport, to know the profile of the CP violation 
near the expanding bubble wall created at the EWPT. In the literatures,
however, some functional forms of the CP violation were assumed without
any reasoning. They must be determined by the dynamics of 
the scalar fields near the EWPT. One may expect that, in the lowest order
of the approximation, spacetime-varying CP violation would be governed
by the classical equations of motion of the gauge-Higgs system, in which
the Higgs potential is replaced with the effective potential at the
transition temperature.
This amounts to find the critical bubble, which would be a good approximation
to an expanding bubble if the EWPT proceeds calmly.\par
In this paper, we shall follow this line to obtain the functional form of
the CP-violating phase in the two-Higgs-doublet model.
In section 2, we derive the equation for the CP-violating phase
assuming that the moduli of the Higgs scalars take the kink shape with the
same width. In section 3, we enumerate possible boundary conditions 
of the phase for various choices of parameters in the potential.
Next we present some numerical solutions and the chiral charge flux
in section 4.
The final section is devoted to discussions.
%
%
%
%\section{The Equation for the Phase}
\section{\protect{\large\bf The Equation for the Phase}}
%\subsection{The Equations of Motion}
\subsection{\protect{\bf The Equations of Motion}}
The system we concern is governed by the lagrangian,
\begin{equation}
 {\cal L} = -{1\over4}F^a_{\mu\nu}F^{a\,\mu\nu}-{1\over4}B_{\mu\nu}B^{\mu\nu}
            +\sum_{i=1,2}\left(D_\mu\Phi_i\right)^\dagger D^\mu\Phi_i
            -V_{eff}(\Phi_1,\Phi_2;T),
	\label{eq:lagrangian}
\end{equation}
where
$$
  D_\mu\Phi_i(x) \equiv (\del_\mu-ig{{\tau^a}\over2}A^a_\mu(x)
                       -i{{g^\prime}\over2}B_\mu(x))\Phi_i(x).
$$
We adopted $V_{eff}$ at $T$ near the transition temperature as the Higgs
potential. We are now interested in the classical solution of the bubble-wall
shape which mediates between the broken and symmetric phases.
If the phase transition proceeds calmly, it will be valid to expect that the
bubble wall grows keeping the profile of the critical bubble, which is
determined by the static equations of motion.
Further, when the bubble is spherically symmetric or is sufficiently
macroscopic so that it is regarded as a planar object, the system is reduced
to an effective one-dimensional one. In general, in $1+1$-dimensional gauge
theories, gauge fields have no dynamical degrees of freedom, that is, they
are pure gauge. Here we assume that the gauge fields are written in the
pure-gauge form:
$$
  ig{{\tau^a}\over2}A^a_\mu(x)=\del_\mu U_2(x)U_2^{-1}(x),\qquad
  i{{g^\prime}\over2}B_\mu(x) =\del_\mu U_1(x)U_1^{-1}(x),
$$
where $U_2$ and $U_1$ are elements of $SU(2)_L$ and $U(1)_Y$, respectively.
Since we can completely gauge away these gauge fields, we only need to
consider the equations of motion for the Higgs fields. Assuming that
$U(1)_{em}$ is not broken anywhere, the vacuum expectation values
(VEVs) of the Higgs fields have the following form:
\begin{equation}
  \langle\Phi_i(x)\rangle 
 =\left(\begin{array}{c} 0 \\
                        {1\over{\sqrt2}}\rho_i(x)\e^{i\theta_i(x)}
        \end{array}\right),\qquad(i=1,2).
	\label{eq:VEV}
\end{equation}
Now the equations of motion are
\begin{eqnarray*}
 -\del^2\rho_i(x)+\rho_i(x)\del_\mu\theta_i(x)\del^\mu\theta_i(x)
 -{{\del V_{eff}}\over{\del\rho_i}} &=& 0,  \\
  \del_\mu\left(\rho^2_i(x)\del^\mu\theta_i(x)\right)
 +{{\del V_{eff}}\over{\del\theta_i}} &=& 0.
\end{eqnarray*}
Along with these equations, we have the ``sourcelessness condition'',
which arises from the requirement for the gauge fields to be pure-gauge type:
$$
  \rho_1^2(x)\del_\mu\theta_1(x) + \rho_2^2(x)\del_\mu\theta_2(x) = 0.
$$
Regarding the bubble wall as a static planar object, these equations are
reduced to (taking $z$ as the coordinate perpendicular to the wall)
\begin{eqnarray}
 {{d^2\rho_i(z)}\over{dz^2}}-\rho_i(z)\left({{d\theta_i(z)}\over{dz}}\right)^2
  -{{\del V_{eff}}\over{\del\rho_i}} &=& 0,
	\label{eq:rho-z}  \\
 {d\over{dz}}\left(\rho_i^2(z){{d\theta_i(z)}\over{dz}}\right)
  -{{\del V_{eff}}\over{\del\theta_i}} &=& 0,
	\label{eq:theta-z}  \\
  \rho_1^2(z){{d\theta_1(z)}\over{dz}}+\rho_2^2(z){{d\theta_2(z)}\over{dz}}
  &=& 0.
	\label{eq:soureless-z}
\end{eqnarray}
For later convenience, let us change the variable $z$ to $y$ defined by
\begin{equation}
 y = {1\over2}\left(1-\tanh(az)\right),    \label{eq:z-y}
\end{equation}
where $a^{-1}$ characterizes the width of the wall.
In terms of this new coordinate, the above equations are written as
\begin{eqnarray}
 4a^2y(1-y){d\over{dy}}\left[y(1-y){{d\rho_i(y)}\over{dy}}\right]
 -4a^2y^2(1-y)^2\rho_i(y)\left({{d\theta_i(y)}\over{dy}}\right)^2
 -{{\del V_{eff}}\over{\del\rho_i}} &=& 0,     \label{eq:rho-y}  \\
 4a^2y(1-y){d\over{dy}}\left[y(1-y)\rho_i^2(y){{d\theta_i(y)}\over{dy}}\right]
 -{{\del V_{eff}}\over{\del\theta_i}} &=& 0,   \label{eq:theta-y}  \\
  \rho_1^2(y){{d\theta_1(y)}\over{dy}}+\rho_2^2(y){{d\theta_2(y)}\over{dy}}
  &=& 0.                                       \label{eq:sourceless-y}
\end{eqnarray}
In order to solve these equations, one must know the explicit form of
$V_{eff}$. Because of the gauge invariance, $V_{eff}$ is a function of
$\theta_1-\theta_2$. From this fact, (\ref{eq:theta-y}) with $i=2$ is
automatically satisfied as long as $\rho_i$ and $\theta_i$ satisfy
(\ref{eq:theta-y}) with $i=1$ and (\ref{eq:sourceless-y}).
%
%
%\subsection{Ansatz for the effective potential}
\subsection{\protect{\bf Ansatz for the effective potential}}
In general, it is difficult to solve the coupled equations (\ref{eq:rho-y})
and (\ref{eq:theta-y}) with the constraint (\ref{eq:sourceless-y}).
When the EWPT proceeds calmly, we expect that the modulus of the Higgs, 
$\rho_i(z)$, has the shape of a kink, and that $\rho_1$ and $\rho_2$ have
the same order of width.\footnote{We implicitly assumed that both the VEVs
acquire nonzero values at about the same temperature.}
Assuming that $\rho_1$ and $\rho_2$ are the kink type of the same width 
but with different amplitudes, the problem is now to solve (\ref{eq:theta-y})
in the background of $\rho_i$. This assumption, in turn, restricts the form
of the effective potential. We shall solve the equations for $\theta_i$
without specifying any model, but with the potential matching this assumption.
\par
We require that (\ref{eq:rho-y}) has the kink-type solutions
in the absence of CP violation ($\theta\equiv\theta_1-\theta_2=0$ or $\pi$);
\begin{equation}
  \rho_i(y) = v_i ( 1 - y ),    \label{eq:rho-kink}
\end{equation}
where
$$
  v_1 = v\cosb,\qquad v_2 = v\sinb.
$$
Then (\ref{eq:rho-y}) is
\begin{equation}
 4a^2v_i y(1-y)(1-2y)+
 \left.{{\del V_{eff}(\rho_1,\rho_2,\theta=0\mbox{ or }\pi))}\over{\del\rho_i}}
 \right|_{\rho_i=v_i(1-y)} = 0.    \label{eq:kink-rho-y}
\end{equation}
For this equation to be satisfied, the form of $V_{eff}$ is somewhat
restricted. Now let us determine $V_{eff}$ in terms of a polynomial
of $\rho_1$ and $\rho_2$.\par
The most general tree-level Higgs potential is given by
\begin{eqnarray}
 V_0&=&
 m_1^2\Phi_1^\dagger\Phi_1 +  m_2^2\Phi_2^\dagger\Phi_2
 +( m_3^2\Phi_1^\dagger\Phi_2 + {\rm h.c.} )                  \nonumber \\
 &+&
 \half\lambda_1(\Phi_1^\dagger\Phi_1)^2+\half\lambda_2(\Phi_2^\dagger\Phi_2)^2
 +\lambda_3(\Phi_1^\dagger\Phi_1)(\Phi_2^\dagger\Phi_2)
 -\lambda_4(\Phi_1^\dagger\Phi_2)(\Phi_2^\dagger\Phi_1)    \label{eq:tree-V}\\
 &-&
 \left\{ \half\lambda_5(\Phi_1^\dagger\Phi_2)^2
 +\bigl[\lambda_6(\Phi_1^\dagger\Phi_1)+\lambda_7(\Phi_2^\dagger\Phi_2)\bigr]
  (\Phi_1^\dagger\Phi_2) + {\rm h.c.} \right\},                 \nonumber
\end{eqnarray}
where $m_1^2, m_2^2, \lambda_1, \lambda_2, \lambda_3, \lambda_4\in{\bf R}$ and
$m_3^2, \lambda_5, \lambda_6, \lambda_7\in{\bf C}$, three of their phases are
independent and yield the explicit CP violation.
When all these parameters are real ({\it i.e.}, no explicit CP violation),
we have
\begin{eqnarray}
 V_0(\rho_1,\rho_2,\theta)&=&
 \half m_1^2\rho_1^2 + \half m_2^2\rho_2^2 + m_3^2\rho_1\rho_2\cos\theta
 +{{\lambda_1}\over8}\rho_1^4 + {{\lambda_2}\over8}\rho_2^4    
       \label{eq:V0-rho-theta}\\
 &+&
 {{\lambda_3-\lambda_4}\over4}\rho_1^2\rho_2^2
 -{{\lambda_5}\over4}\rho_1^2\rho_2^2\cos(2\theta)
 -{{\lambda_6}\over2}\rho_1^3\rho_2\cos\theta
 -{{\lambda_7}\over2}\rho_1\rho_2^3\cos\theta.   \nonumber
\end{eqnarray}
In the following we shall examine the spontaneously CP-violating and
CP-conserving cases.
The CP-conserving case is realized if $\theta=0$ or $\pi$.
It is sufficient to consider the former, since the latter is 
obtained by redefining one of the scalars.
Without any CP violation, we have, in terms of $\rho_i$,
\begin{eqnarray}
 V_0&=&
 \half m_1^2\rho_1^2 + \half m_2^2\rho_2^2 + m_3^2\rho_1\rho_2
 +{{\lambda_1}\over8}\rho_1^4 + {{\lambda_2}\over8}\rho_2^4    \nonumber\\
 &+&
 {{\tilde\lambda_3}\over4}\rho_1^2\rho_2^2
 -\half(\lambda_6\rho_1^2+\lambda_7\rho_2^2)\rho_1\rho_2,   \label{eq:V0-rho}
\end{eqnarray}
where $\tilde\lambda_3=\lambda_3-\lambda_4-\lambda_5$.\par
In order to have kink solutions for $\rho_i$, we need $\rho^3$-terms with
negative coefficients, which are expected to arise at finite temperature.
\footnote{Although we do not use the high-temperature expansion,
these terms may be considered to represent the effect which causes the
first-order phase transition.}
Hence we adopt the following ansatz for the effective potential:
\begin{equation}
  V_{eff}(\rho_1,\rho_2,\theta=0)
 =V_0(\rho_1,\rho_2, 0) 
  -\left( A\rho_1^3 + B\rho_1^2\rho_2 + C\rho_1\rho_2^2 + D\rho_2^3\right).
     \label{eq:ansatz-V}
\end{equation}
Since we expect this to represent the effective potential with first-order
phase transition, the origin $(\rho_1,\rho_2)=(0,0)$ and the point
$(\rho_1,\rho_2)=(v\cosb,v\sinb)$ must be local minima.
This condition amounts to
\begin{equation}
 {\rm det}\left({{\del^2 V_{eff}}\over{\del\rho_i\del\rho_j}}\right) >0,
 \qquad{{\del^2 V_{eff}}\over{\del\rho_1^2}}>0\mbox{ or }
       {{\del^2 V_{eff}}\over{\del\rho_2^2}}>0,
      \label{eq:cond-min}
\end{equation}
at each point. At $(0,0)$, this reduces to
\begin{equation}
  m_1^2 m_2^2-m_3^4 > 0,\quad\mbox{and $m_1^2>0$ or $m_2^2>0$}.
      \label{eq:cond-min-0}
\end{equation}
Requiring that (\ref{eq:kink-rho-y}) is satisfied with this potential, 
some of the parameters are expressed by the others.
Now we assume $\cosb\sinb\not=0$.
This condition would be necessary to give masses to up- and down-type
quarks when they couple to different Higgs doublets, as in the MSSM.
Then the effective potential in the absence of CP violation is
\begin{eqnarray}
 V_{eff}(\rho_1,\rho_2,0)&=&
 (2a^2-\half m_3^2\tan\beta)\rho_1^2+(2a^2-\half m_3^2\cot\beta)\rho_2^2
 + m_3^2\rho_1\rho_2                     \nonumber \\
 &-&
  \left\{ A\rho_1^3
 +\left[-2A\cot\beta+D\tan^2\beta+{{4a^2}\over{v\sinb}}(3-{1\over{\cos^2\beta}})
       \right]\rho_1^2\rho_2 \right.             \nonumber \\
 & &
  \left.
 +\left[A\cot^2\beta-2D\tan\beta+{{4a^2}\over{v\cosb}}(3-{1\over{\sin^2\beta}})
       \right]\rho_1\rho_2^2 + D\rho_2^3 \right\}      \nonumber \\
 &+&
  {{\lambda_1}\over8}\rho_1^4+{{\lambda_2}\over8}\rho_2^4
 +{{\tilde\lambda_3}\over4}\rho_1^2\rho_2^2            \label{eq:Veff-rho}\\
 &-&
  {1\over8}\left\{
  \left[{3\over2}\lambda_1\cot\beta-{{\lambda_2}\over2}\tan^3\beta
 +\tilde\lambda_3\tan\beta
 -{{8a^2}\over{v^2\sinb\cosb}}(4-{1\over{\cos^2\beta}})\right]\rho_1^3\rho_2
   \right.    \nonumber \\
 & &
  \left.
 +\left[-{{\lambda_1}\over2}\cot^3\beta + {3\over2}\lambda_2\tan\beta
 +\tilde\lambda_3\cot\beta
 -{{8a^2}\over{v^2\sinb\cosb}}(4-{1\over{\sin^2\beta}})\right]\rho_1\rho_2^3
 \right\},         \nonumber
\end{eqnarray}
and the condition (\ref{eq:cond-min-0}) is written as
\begin{equation}
  4a^2 > {{m_3^2}\over{\sinb\cosb}}.  \label{eq:cond-min-01}
\end{equation}
Note that although we use the same notations for the parameters in $V_{eff}$
as those in $V_0$, their meanings are different.
Those in $V_{eff}$ contain radiative as well as finite-temperature 
corrections near $T_C$.\par
Now let us introduce the CP-violating phase into (\ref{eq:Veff-rho})
in a gauge-invariant manner.
Comparing (\ref{eq:V0-rho-theta}) and (\ref{eq:V0-rho}) suggests that 
the phase $\theta$ is introduced in $V_{eff}$ of (\ref{eq:Veff-rho}) as
\begin{eqnarray*}
  \rho_1\rho_2 &\rightarrow& \rho_1\rho_2\cos\theta,  \\
  \lambda_5\rho_1^2\rho_2^2 &\rightarrow&
     \lambda_5\rho_1^2\rho_2^2\cos(2\theta),   \\
  \rho_1^3\rho_2 &\rightarrow& \rho_1^3\rho_2\cos\theta,  \\
  \rho_1\rho_2^3 &\rightarrow& \rho_1\rho_2^3\cos\theta.
\end{eqnarray*}
On the other hand, we have no principle to determine $\theta$-dependence
of $\rho^3$-terms.
Here we investigate the two possibilities:
\begin{itemize}
 \item  no $\theta$-dependence in the $\rho^3$-terms,
 \item  $\rho_1^2\rho_2\rightarrow\rho_1^2\rho_2\cos\theta$,
        $\rho_1\rho_2^2\rightarrow\rho_1\rho_2^2\cos\theta$.
\end{itemize}
Hence our ansatz for the effective potential is
\begin{eqnarray}
& & V_{eff}(\rho_1,\rho_2,\theta)   \nonumber \\
&=&
 (2a^2-\half m_3^2\tan\beta)\rho_1^2+(2a^2-\half m_3^2\cot\beta)\rho_2^2
 + m_3^2\rho_1\rho_2\cos\theta           \nonumber \\
 &-&
  \left\{ A\rho_1^3
 +\left[-2A\cot\beta+D\tan^2\beta+{{4a^2}\over{v\sinb}}(3-{1\over{\cos^2\beta}})
       \right]\rho_1^2\rho_2(\cos\theta) \right.    \nonumber \\
 & &
  \left.
 +\left[A\cot^2\beta-2D\tan\beta+{{4a^2}\over{v\cosb}}(3-{1\over{\sin^2\beta}})
       \right]\rho_1\rho_2^2(\cos\theta) + D\rho_2^3 \right\}\nonumber \\
 &+&
  {{\lambda_1}\over8}\rho_1^4+{{\lambda_2}\over8}\rho_2^4
 +{{\lambda_3-\lambda_4}\over4}\rho_1^2\rho_2^2
 -{{\lambda_5}\over4}\rho_1^2\rho_2^2\cos(2\theta)  \label{eq:full-Veff}\\
 &-&
  {1\over8}\left\{
  \left[{3\over2}\lambda_1\cot\beta-{{\lambda_2}\over2}\tan^3\beta
 +\tilde\lambda_3\tan\beta
 -{{8a^2}\over{v^2\sinb\cosb}}(4-{1\over{\cos^2\beta}})\right]\rho_1^3\rho_2
   \right.    \nonumber \\
 & &
  \left.
 +\left[-{{\lambda_1}\over2}\cot^3\beta + {3\over2}\lambda_2\tan\beta
 +\tilde\lambda_3\cot\beta
 -{{8a^2}\over{v^2\sinb\cosb}}(4-{1\over{\sin^2\beta}})\right]\rho_1\rho_2^3
 \right\}\cos\theta.         \nonumber
\end{eqnarray}
Here $\cos\theta$ in the $\rho^3$-terms is unity in the case of the first
possibility.
%
%\subsection{Equations for the phases}
\subsection{\protect{\bf Equations for the phases}}
Once $\theta$-dependence of $V_{eff}$ is determined, one can derive
the equations for $\theta_i$ from (\ref{eq:theta-y}).
In our case, the sourcelessness condition (\ref{eq:sourceless-y})
is reduced to
\begin{equation}
 \cos^2\beta{{d\theta_1}\over{dy}}+\sin^2\beta{{d\theta_2}\over{dy}}=0.
  \label{eq:sourceless-kink}
\end{equation}
As noted above, we only need to solve the equation for $\theta_1$, as long as
$\theta_1$ and $\theta_2$ satisfy the sourcelessness condition.
Since, from this condition, $\theta_2(y)$ is written as
$$
  \theta_2(y) = -\theta_1(y)\cot^2\beta + \mbox{const.},
$$
we have
$$
  \theta(y) = {1\over{\sin^2\beta}}(\theta_1(y)+\mbox{const.}).
$$
Noting that the derivative terms in the equation for $\theta_1$ are invariant
under the shift of $\theta_1$, the constant in the r.h.s. of the above equation
can be ignored.
Putting
\begin{equation}
  \theta(y)={1\over{\sin^2\beta}}\theta_1(y),   \label{eq:def-theta}
\end{equation}
we have
\begin{eqnarray}
 & & y^2(1-y)^2{{d^2\theta(y)}\over{dy^2}}
 +y(1-y)(1-4y){{d\theta(y)}\over{dy}}        \nonumber \\
 &=&
 {1\over{4a^2\sin^2\beta\cos^2\beta}}\left[
 -m_3^2\sinb\cosb 
 -\left( A\cos^3\beta+D\sin^3\beta-{{4a^2}\over v} \right) v(1-y)
        \right.        \nonumber\\
 & &\qquad\left.
 +{{v^2}\over8}(\lambda_1\cos^4\beta+\lambda_2\sin^4\beta
  +2\tilde\lambda_3\sin^2\beta\cos^2\beta-{{16a^2}\over{v^2}})(1-y)^2\right]
 \sin\theta(y)     \nonumber\\
 & &\qquad
 +{{\lambda_5 v^2}\over{4a^2}}(1-y)^2\sin\theta(y)\cos\theta(y).
   \label{eq:eq-theta-row}
\end{eqnarray}
\par
For later convenience, we denote the coefficients in the above equation
as
\begin{eqnarray}
  b&\equiv& -{{m_3^2}\over{4a^2\sinb\cosb}},   \nonumber\\
  c&\equiv& {{v^2}\over{32a^2}}(\lambda_1\cot^2\beta+\lambda_2\tan^2\beta
            +2\tilde\lambda_3) - {1\over{2\sin^2\beta\cos^2\beta}}
      \nonumber \\
   &=&  {{v^2}\over{8a^2}}(\lambda_6\cot\beta + \lambda_7\tan\beta), 
                  \label{eq:def-bcd}\\
  d&\equiv& {{\lambda_5 v^2}\over{4a^2}}.   \nonumber   \\
  e&\equiv& {v\over{4a^2\sin^2\beta\cos^2\beta}}
             \left( A\cos^3\beta+D\sin^3\beta-{{4a^2}\over v} \right)
                   \nonumber\\
   &=& -{v\over{4a^2}}\left({B\over\sinb}+{C\over\cosb}\right). \nonumber
\end{eqnarray}
If there is no $\theta$-dependence in the $\rho^3$-terms
in the $V_{eff}$, $e=0$.
Because of the condition (\ref{eq:cond-min-01}),
\begin{equation}
   b > -1.                     \label{eq:cond-min-02}
\end{equation}
On the other hand, at $(\rho_1,\rho_2)=(v\cosb,v\sinb)$
\begin{eqnarray*}
 {{\del^2 V_{eff}}\over{\del\rho_1^2}}&=&
 m_1^2-(8a^2e-12a^2c+\tilde\lambda_3v^2)\sin^2\beta,  \\
 {{\del^2 V_{eff}}\over{\del\rho_2^2}}&=&
 m_2^2-(8a^2e-12a^2c+\tilde\lambda_3v^2)\cos^2\beta,  \\
 {{\del^2 V_{eff}}\over{\del\rho_1\del\rho_2}}&=&
 m_3^2+(8a^2e-12a^2c+\tilde\lambda_3v^2)\sinb\cosb,
\end{eqnarray*}
so that
$$
 {\rm det}\left({{\del^2 V_{eff}}\over{\del\rho_i\del\rho_j}}\right)
=16a^4\left(1+b-2e+3c-{{\tilde\lambda_3v^2}\over{4a^2}}\right).
$$
Thus the condition (\ref{eq:cond-min}) is satisfied if
\begin{equation}
 b-2e+3c > -1 + {{\tilde\lambda_3v^2}\over{4a^2}}.  \label{eq:cond-min-1}
\end{equation}
Now (\ref{eq:eq-theta-row}) is written as
\begin{equation}
  y^2(1-y)^2{{d^2\theta(y)}\over{dy^2}}+y(1-y)(1-4y){{d\theta(y)}\over{dy}}
 =[b+c(1-y)^2-e(1-y)]\sin\theta(y) + {d\over2}(1-y)^2\sin(2\theta(y)).
	\label{eq:eq-theta}
\end{equation}
This is the equation that we shall examine in detail.
One sees that $\theta(y)=n\pi$ with $n\in{\bf Z}$ is the trivial solution.
This equation is invariant under $\theta(y)\mapsto-\theta(y)$. This is because
we have no explicit CP-violating terms in the potential.
%Further (\ref{eq:eq-theta}) has the symmetry under
%\begin{eqnarray}
%    \theta(y) &\mapsto& \theta(y)\pm\pi,     \label{eq:symmetry-b}  \\
%     (b,c,d)   &\mapsto& (-b,-c,d).           \nonumber
%\end{eqnarray}
\par
Before closing this section, we comment on the ansatz adopted.
Although the kinks (\ref{eq:rho-kink}) are solutions of (\ref{eq:kink-rho-y})
with the potential (\ref{eq:Veff-rho}), they are no longer solutions of the
coupled equations for $\rho_i$ and $\theta_i$ with the potential
(\ref{eq:full-Veff}). So that solutions to (\ref{eq:eq-theta-row}) will not
be true solutions of the coupled equations.
Further some of the parameters may be restricted to give a finite-energy
solution for $\theta$.
We, however, expect that our solutions are not so different from the
true solutions. This is because, as long as the $\rho$'s have the kink shape,
$\rho$'s and $\theta$'s continuously rearrange themselves so as to take 
the minimal-energy configuration starting from our solutions.
At the final stage, we should check that the physical quantities,
such as the generated baryon number, are insensitive to small perturbations
in the parameters of the potential.
%
%
%
%\section{Asymptotic Behaviors of $\theta$}
\section{\protect{\large\bf Asymptotic Behaviors of $\theta$}}
Before solving (\ref{eq:eq-theta}), we investigate the asymptotic behaviors
of the solutions. This will help us to find numerical solutions.
Among possible solutions, we are concerned in those with finite energy density.
When all the gauge fields are gauged away, the classical energy of the
bubble wall is given only by the contribution of the scalars:
$$
 E=\int d^3{\bf x}\left\{
   \sum_{i=1,2}\left[ \dot\Phi_i^\dagger(x)\dot\Phi_i(x)
     + \nabla\Phi_i^\dagger(x)\cdot\nabla\Phi_i(x) \right]
   + V_{eff}(\Phi_1,\Phi_2;T) \right\}.
$$
For a static and planar bubble wall, the energy density per unit area is,
in terms of $\rho_i$ and $\theta_i$,
\begin{eqnarray}
 {\cal E}&=&\int_{-\infty}^\infty dz \left\{
  \half\sum_{i=1,2}\left[\left({{d\rho_i}\over{dz}}\right)^2
    +\rho_i^2 \left({{d\theta_i}\over{dz}}\right)^2 \right]
   + V_{eff}(\rho_1,\rho_2,\theta) \right\}      \nonumber\\
 &=&
  \int_0^1 dy \left\{
  ay(1-y)\sum_{i=1,2}\left[\left({{d\rho_i}\over{dy}}\right)^2
           +\rho_i^2 \left({{d\theta_i}\over{dy}}\right)^2 \right] 
              \right.     \label{eq:energy-y}\\
 & &\left.\qquad\qquad 
  +{1\over{2ay(1-y)}}V_{eff}(\rho_1,\rho_2,\theta) \right\}.      
           \nonumber
\end{eqnarray}
In the case of the kink-type wall treated in the previous section,
for the energy density to be finite, we must have
\begin{eqnarray}
 \theta_i^\prime(y)&\sim& y^\alpha,\quad\mbox{with $\alpha>-1$}
         \quad\mbox{for $y\sim0$},   \label{eq:E-finite-0}\\
 \theta_i^\prime(y)&\sim& (1-y)^\beta,\quad\mbox{with $\beta>-2$}
         \quad\mbox{for $y\sim1$}.   \label{eq:E-finite-1}
\end{eqnarray}
%
%
%\subsection{Asymptotic behavior in the broken phase}
\subsection{\protect{\bf Asymptotic behavior in the broken phase}}
The potential with the kink-type profile (\ref{eq:rho-kink}) for $\rho_i$
is written as
\begin{eqnarray}
  V_{eff}(\rho_1,\rho_2,\theta) 
&=&
  2a^2v^2y^2(1-y)^2          \nonumber\\
& &
  +4a^2v^2\sin^2\beta\cos^2\beta(1-y)^2\left\{
   [b+c(1-y)^2-e(1-y)](1-\cos\theta) \right.    \nonumber\\
& &
  \left. + {d\over4}(1-y)^2(1-\cos2\theta) \right\}. \label{eq:Veff-kink}
\end{eqnarray}
Although in the symmetric phase the value of $\theta$ is not
determined because of $V_{eff}(y=1)=0$, it will be allowed to take some
specific values as shown later. On the other hand, in the broken phase
($y=0$),
\begin{equation}
 V_{eff}(\theta_0) =
 4a^2v^2\sin^2\beta\cos^2\beta\left[(b+c-e)(1-\cos\theta_0)
 +{d\over4}(1-\cos2\theta_0) \right] \label{eq:Veff-Delta-theta-1},
\end{equation}
where $\theta_0=\theta(0)$.
If $d=0$ ($\lambda_5=0$), $\theta_0=2n\pi((2n+1)\pi)$ for positive
(negative) $b+c-e$, {\it i.e.}, no CP violation.
When $d\not=0$, (\ref{eq:Veff-Delta-theta-1}) is written as
\begin{equation}
 V_{eff}(\theta_0)
=-2a^2 v^2 d\sin^2\beta\cos^2\beta
  \left(\cos\theta_0+{{b+c-e}\over d}\right)^2
  +\mbox{$\theta_0$-indep. terms}.       \label{eq:Veff-Delta-theta-2}
\end{equation}
This implies that
\begin{eqnarray*}
  \cos\theta_0&=&{{b+c-e}\over{-d}},\qquad\mbox{for $d<0$ and $\abs{b+c-e}<-d$}, \\
  \theta_0 &=& 2n\pi, \qquad\mbox{for $0<-d<b+c-e$, or $d>0$ and $b+c-e>0$}, \\
  \theta_0 &=&(2n+1)\pi,\qquad\mbox{for $b+c-e<d<0$, or $d>0$ and $b+c-e<0$}.
\end{eqnarray*}
\par
Suppose that $\theta(y)$ is expanded as
\begin{equation}
 \theta(y)=\theta_0+y^\nu\sum_{n=0}^\infty a_n y^n \qquad(\nu>0,a_0\not=0)
   \label{exp-theta-0}
\end{equation}
at $y\sim0$.
The constraint $\nu>0$ matches the condition (\ref{eq:E-finite-0}).
When inserted in (\ref{eq:eq-theta}), this yields
\begin{eqnarray}
 & & 
 y^\nu\left\{ \nu^2a_0 +\left[(\nu+1)^2a_1-\nu(2\nu+3)a_0\right]y \right.
       \nonumber\\
 & &
 +\sum_{n=2}^\infty\left[(n+\nu)^2a_n \right.      \nonumber\\
 & &\left.\left.
 \qquad - (n+\nu-1)(2n+2\nu+1)a_{n-1}
   + (n+\nu-2)(n+\nu+1)a_{n-2}\right] y^n \right\}   \nonumber\\
 &=&
 y^\nu\left\{ [b+c(1-y)^2-e(1-y)]\cos\theta_0 \left[\sum_{n=0}^\infty a_n y^n
   -{1\over{3!}}y^{2\nu}\bigl(\sum_{n=0}^\infty a_n y^n\bigr)^3
   + \cdots \right]   \right.       \label{eq:exp-broken}\\
 & &\left.\quad
   +{d\over2}(1-y)^2\cos(2\theta_0) \left[ 2\sum_{n=0}^\infty a_n y^n
   -{{2^3}\over{3!}}y^{2\nu}\bigl(\sum_{n=0}^\infty a_n y^n\bigr)^3
   + \cdots \right] \right\}          \nonumber\\
 &+&
  [b+c(1-y)^2-e(1-y)]\sin\theta_0 \left[ 1
   -{1\over{2!}}y^{2\nu}\bigl(\sum_{n=0}^\infty a_n y^n\bigr)^2+\cdots\right]
           \nonumber\\
 & &
  +{d\over2}(1-y)^2\sin(2\theta_0) \left[ 1
  -{{2^2}\over{2!}}y^{2\nu}\bigl(\sum_{n=0}^\infty a_n y^n\bigr)^2
   + \cdots \right].       \nonumber
\end{eqnarray}
In order to have a nontrivial solution, $2\nu\in{\bf Z}$.\par
When $\theta_0=n\pi$ with $n\in{\bf Z}$, the lowest order terms give
\begin{equation}
  (b+c-e)\cos\theta_0 + d = \nu^2.    \label{eq:yI0-0}
\end{equation}
The higher order terms of $y$ will relate $a_n$ with $n\ge1$ to the lower
coefficients in a nonlinear way.\par
When $\theta_0\not=n\pi$, $\nu$ must be an integer for a nontrivial solution
to exist. Then the lowest order terms yield
\begin{equation}
  b+c-e+d\cos\theta_0 = 0.     \label{eq:yII0-0}
\end{equation}
For this to be satisfied,
\begin{equation}
    \abs{b+c-e}\le\abs{d}.     \label{eq:CP-viol-0}
\end{equation}
As we saw above, this $\theta_0$ is energetically realized when
$\abs{b+c-e}<-d$ with $d<0$.
For $\nu=1$, $O(y)$-terms of (\ref{eq:exp-broken}) lead to
$$
  a_0 = [(b+c-e)\cos\theta_0+d\cos(2\theta_0)]a_0
        -(2c-e)\sin\theta_0 - d\sin(2\theta_0),
$$
which means, by use of (\ref{eq:yII0-0}),
\begin{equation}
  a_0={{(2b-e)\sin\theta_0}\over{1+d\sin^2\theta_0}}.  
       \label{eq:a0-broken-1}
\end{equation}
The higher order terms give relations among $a_n$ with $n\ge1$ and the lower
coefficients. For $\nu\ge 2$, $O(y)$-terms of (\ref{eq:exp-broken}) yield
$2b-e=0$. When $\nu=2$, $O(y^2)$-terms of (\ref{eq:exp-broken}) give
\begin{equation}
  a_0=-{{(b-e)\sin\theta_0}\over{4+d\sin^2\theta_0}}.  
       \label{eq:a0-broken-2}
\end{equation}
On the other hand, $O(y^2)$-terms for $\nu\ge 3$ lead to $b=e$, which means
$b=e=0$.
We discard this case because of two reasons:
One is because nonvanishing $b\propto m_3^2$ is needed to violate
CP spontaneously when $\lambda_{5,6,7}=0$ as in the case of MSSM.
The other is because $m_3^2$ in $V_{eff}$ is induced in the presence of
tree-level $\lambda_{5,6,7}$ so that to have $m_3^2=0$ is unnatural.
Hence in this case, only $\nu=1$ and $\nu=2$ are allowed.
Note that since $a_0=\theta'(0)$, the initial conditions for
(\ref{eq:eq-theta}) are completely fixed by the parameters of
the potential.\par
To summarize, the possible boundary values of $\theta_0$ are as follows:
\begin{enumerate}
 \item[(a)] $\theta_0=2n\pi$, if $b+c-e>-d>0$, or $d\ge 0$ and $b+c-e>0$.
  The parameters must satisfy $b+c-e+d=\nu^2$ with $2\nu\in{\bf Z}$.  
 \item[(b)] $\theta_0=(2n+1)\pi$, if $b+c-e<d<0$, or $d\ge 0$ and $b+c-e<0$.
  The parameters must satisfy $-(b+c-e)+d=\nu^2$ with $2\nu\in{\bf Z}$.
 \item[(c)] $\cos\theta_0=-(b+c-e)/d$, if $\abs{b+c-e}<-d$ and $d<0$,
  for $\nu=1\mbox{ or }2$.
\end{enumerate}
Among these, we do not need to consider the case (b), as discussed in
section 2.
%
%
%\subsection{Asymptotic behavior in the symmetric phase}
\subsection{\protect{\bf Asymptotic behavior in the symmetric phase}}
Define $\z=1-y$ and suppose that $\theta(\z)$ is expanded as
\begin{equation}
 \theta(\z)=\theta_1+\z^\mu\sum_{n=0}^\infty b_n \z^n \qquad(\mu>0,b_0\not=0)
   \label{exp-theta-1}
\end{equation}
at $y\sim1$ ($\z\sim0$).
In terms of $\z$, (\ref{eq:eq-theta}) is written as
\begin{equation}
  \z^2(1-\z)^2{{d^2\theta(\z)}\over{d\z^2}}
 +\z(1-\z)(3-4\z){{d\theta(\z)}\over{d\z}}
 =(b+c\z^2-e\z)\sin\theta(\z) + {d\over2}\z^2\sin(2\theta(\z)). \label{eq:eq-theta-z}
\end{equation}
The condition $2\mu\in{\bf Z}$ is required to have a nontrivial solution.
The lowest order terms lead to
\begin{equation}
  b\cos\theta_1 = \mu(\mu+2)    \label{eq:zi0-0}
\end{equation}
when $\theta_1\equiv\theta(1)=n\pi$, and
\begin{equation}
  b = 0     \label{eq:zii0-0}
\end{equation}
and $\mu\in{\bf Z}$ when $\theta_1\not=n\pi$.
The higher order terms will give relations among the expansion coefficients.
One may think that it is meaningless to ask the value of $\theta_1$, since
in the symmetric phase the Higgs fields vanish so that CP is never 
violated in the Higgs sector. But what is important for 
the baryogenesis is how the phases of them tend to some value as the Higgs
fields disappear. We shall not discuss the case of $b=0$, {\it i.e.},
$m_3^2=0$ for the reason stated above. Besides this, the first-order EWPT is
realized when $b>-1$ (see (\ref{eq:cond-min-02})).
Hence the possible boundary values $\theta_1$ are;
\begin{equation}
 \theta_1=2n\pi,\qquad\mbox{$b=\mu(\mu+2)$ with $2\mu\in{\bf Z}$}.
      \label{eq:theta-1}  
\end{equation}
%
%
%
%\section{Numerical Analysis}
\section{\protect{\large\bf Numerical Analysis}}
As we saw in the previous section, we are concerned with solutions 
satisfying the boundary conditions either (a) or (c), and 
(\ref{eq:theta-1}), according to the parameters in the effective potential.
We present numerical solutions satisfying each set of boundary conditions.
One corresponds to the case in which CP is spontaneously violated in
the broken phase, while the other has no CP violation in the broken phase.
To find such solutions, we performed numerical analysis using the
relaxation, as well as the shooting, algorithms, with the parameters
taken to satisfy the conditions (\ref{eq:cond-min-02}) and 
(\ref{eq:cond-min-1}).
We also evaluated the chiral charge flux with use of these numerical 
solutions, which is the basic quantity to generate the baryon number 
in the charge transport scenario\cite{NKC}.
%
%
%\subsection{Solutions with spontaneous CP violation in the broken phase}
\subsection{\protect{\bf Solutions with spontaneous CP violation
in the broken phase}}
Although one may think that nonzero $\theta_0$ is induced in some model 
at finite temperature\cite{Comelli}, we treat it as an input parameter.
Once $\theta_0$ is fixed, we can choose three of the four parameters
$(b,c,d,e)$ (see (c) in section 3). Because of (\ref{eq:theta-1}), 
$b=5/4,3,21/4,8\ldots$.\par
As an example, we show the solution for the case of $\theta_0=0.002$
and $(b,c,e) = (3,7,7)$ in Fig.~1. 
This suggests that the real part of the VEV has the kink shape and 
the imaginary part can be regarded as a perturbation to it.
Then the effects of the CP violation on the fermions scattered off the wall,
which interact with the Higgs through the Yukawa coupling,
can be treated by the perturbative method developed in \cite{FKOTTa}.
Recalling that $\rho_i\propto 1-y$ for the kink-type profile,
the behavior of $\theta(y)\sim\theta_0(1-y)$ with small $\theta_0$ in
Fig.~1 implies that the imaginary part of the VEV is approximately
proportional to the square of the real part, just as in the case studied in
\cite{FKOTTb}.\footnote{To $O(\theta_0)$, $\theta(y)=\theta_0(1-y)^\mu$
with $2\mu\in{\bf Z}$ is an approximate solution to (\ref{eq:eq-theta})
for $b=\mu(\mu+2)$, $e=\mu(2\mu+5)$ and $d=\mu(\mu+3)-c<0$.}
We shall not repeat the calculation of the chiral charge
flux and note that $\Delta\theta$ in \cite{FKOTTb} should be replaced
with $\theta_0$ here.\par
Besides this solution, we also found a solution with $\theta_0=1$ for
$(b,c,e) = (3,-1,0)$, shown in Fig.~2. 
Unlike the above solution, it no longer has
a linear shape and the perturbative method is not applicable.
Such a solution would be realized when CP violation in the broken phase
is enhanced by finite temperature effects and almost disappears at zero
temperature as the case studied by Comelli, {\it et al.}\cite{Comelli}.
%
%
%\subsection{Solutions without CP violation in the broken phase}
\subsection{\protect{\bf Solutions without CP violation in the broken phase}}
In this case the parameters must satisfy (a).
Then $(\rho_1,\rho_2,\theta)=(v\cosb(1-y),v\sinb(1-y),2n\pi)$ is an exact
solution to the full equations of motion (\ref{eq:rho-y}) and 
(\ref{eq:theta-y}), which we refer to as the trivial solution.
If there exists a nontrivial solution to (\ref{eq:eq-theta}),
it will open an interesting possibility that nonzero baryon asymmetry can be
generated even in a CP-conserving theory in the broken phase.\par
We found such a nontrivial solution for $(b,c,d,e)=(3,12.2,-2,12.2)$,
whose profile is plotted in Fig.~3. From this, the real and imaginary
parts of the VEV are obtained and shown in Fig.~4, which suggest that
the perturbative method is applicable.
We calculated the chiral charge flux for various choices of the fermion
mass $m_0$ and wall width. The results are summarized as the contour plot in
Fig.~5. This shows that the chiral charge flux is comparable to those
studied in \cite{FKOTTb}, so that it could generate sufficient baryon
number for the thin wall case, taking into account the enhancement of forward
scattering.\par
The energy density per unit area of this solution is
\begin{equation}
  \Delta{\cal E} = {\cal E} - \left.{\cal E}\right|_{\theta=0}
                 = - 2.056\times 10^{-3}\,av^2\sin^2\beta\cos^2\beta,
           \label{eq:energy-density-diff}
\end{equation}
where ${\cal E}$ is defined by (\ref{eq:energy-y}) and
$\left.{\cal E}\right|_{\theta=0}=av^2/3$.
Since this solution is {\it not} a solution to the full equations of motion,
the energy of the true solution may be lower than this value.
We found that the solution is stable under perturbation of the
parameters, so that we expect that this type of solutions will exist
even if we do not impose the kink-type profile for $\rho_i$.\par
For such a nontrivial solution to exist, the CP-violating state must
be favored in the intermediate region between the broken and symmetric
phases. From (\ref{eq:Veff-kink}), the effective potential along the kink
is 
\begin{eqnarray}
  V_{eff}(\z,\theta) 
&=&
  2a^2v^2\left\{ \z^2(1-\z)^2   \right.       \label{eq:Veff-kink-z}\\
& &\left.
  -d\z^4\sin^2\beta\cos^2\beta\left[
   \left(\cos\theta+{{b+c\z^2-e\z}\over{d\z^2}}\right)^2
  -\left(1+{{b+c\z^2-e\z}\over{d\z^2}}\right)^2 \right] \right\},
                     \nonumber
\end{eqnarray}
where $\z=1-y$. Just as the boundary value $\theta_0$, $\theta(y)$ can
take a value other than $n\pi$, when $d<0$ and there is a region in 
$0<\z<1$ such that $\abs{(b+c\z^2-e\z)/(d\z^2)}<1$.
Further minimization with respect to $\z$ in such a region will give the
local minimum of $V_{eff}$. Since this new minimum appears on coupling the phase
$\theta$, the expected structure of $V_{eff}(\rho_1,\rho_2,\theta)$ is
somehow modified compared with $V_{eff}(\rho_1,\rho_2,0)$, which has
degenerate local minima at $(\rho_1,\rho_2,\theta)=(0,0,0)$ and
$(v\cosb,v\sinb,0)$ reflecting the first-order nature of the EWPT.
Hence we must require that the new minimum with $\theta\not=n\pi$ should not 
be so deep, otherwise it drastically changes the dynamics of the EWPT.
Here we shall not investigate this condition further, but only assure that
the new minimum is not lower than $(0,0,0)$. 
For the above set of the parameters and $\beta=\pi/4$, for which 
the negative contribution is maximal, $V_{eff}$ as a function of 
$(y,\theta)$ is shown in Fig.~6.
The new minimum of $V_{eff}$ is about $0.0069a^2v^2$ and the height of the
saddle point between it and the origin is about $0.038a^2v^2$,
which is much lower than the maximum $0.125a^2v^2$ along $\theta=0$.
Since the barrier between the new minimum and the origin is not so high,
we expect that the nature of the EWPT is not essentially altered.
The contour plot of $V_{eff}$ also shows that the numerical solution for
$\theta(y)$ goes around the maximum at $(1/2,0)$.
The value of $V_{eff}$ where the maximum of $\theta(y)$ is reached
is about $0.116a^2v^2$, which is just below the CP-conserving maximum.
This fact suggests that this type of solution may exist irrespective of
the depth of the new local minimum.
%
%
%
%\section{Discussions}
\section{\protect{\large\bf Discussions}}
Based on a rather general effective potential in the two-Higgs-doublet model
at the transition temperature, which exhibits first-order EWPT,
we have classified possible finite-energy bubble wall solutions with 
CP violation.
We have found two types of numerical solutions, which are characterized by
the CP-violating angle in the broken phase, $\theta_0$.\par
One of them is the solution in the case where CP is spontaneously violated
in the broken phase and will be the lowest-energy solution in that case.
It smoothly mediates between the CP-violating vacuum in 
the broken phase and the CP-conserving one in the symmetric phase.
For sufficiently small $\theta_0$, the profile of the bubble wall
is similar to that used in the literatures to estimate generated baryon
numbers, while for $\theta_0=O(1)$, $\theta(y)$ is no longer proportional
to the kink so that the previous estimations of generated baryon number with
large $\theta$, by use of some presumed profile, should be revised.\par
The other connects the CP-conserving vacua in both phases.
When CP is not violated in both vacua, there always exists the trivial
solution with $\theta=0$ all along. The new solution has nonvanishing
$\theta$ within the bubble wall and has lower energy than the trivial one,
as shown in (\ref{eq:energy-density-diff}).
For the critical bubble of radius $R_C$, this CP-violating bubble
will be nucleated with more probability than the trivial one by the factor
\begin{equation}
  {\rm exp}\left( -{{4\pi R_C^2\Delta{\cal E}}\over{T_C}} \right).
        \label{eq:prob-CP-viol-bubble}
\end{equation}
According to the estimation in the massless two-Higgs-doublet 
model\cite{FKT}, the radius of the critical bubble is given by
$\sqrt{3F_C/(4\pi av^2)}$, where $F_C$ is the free energy of the
critical bubble and is found to be about $145T$. Then the exponent in
(\ref{eq:prob-CP-viol-bubble}) is $0.89\sin^2\beta\cos^2\beta$, irrespective
of the wall width.
This solution will provide a new possibility to generate baryon number
even within the framework of a CP-conserving theory.\footnote{
Strictly speaking, one needs a CP-odd term in the energy to have 
net baryon number by solving the degeneracy between the solutions with
$\theta(y)$ and $-\theta(y)$.}\par
We estimated the generated baryon number with the use of each solution,
based on the charge transport mechanism. With the enhancement of forward
scattering, we could have sufficient baryon number for thin wall case.
Although we did not work out on the spontaneous scenario, we expect that
sufficient baryon number could be obtained if the effects of diffusion
are considered\cite{diffusion}.\par
We found that each type of the solutions are stable under perturbation of the
parameters in $V_{eff}$, as long as they satisfy the required conditions.
Hence these kinds of solutions will exist in wider class of models.
It will be interesting to find a realistic model which admits such a 
solution.\par
When we almost completed this work, we noticed a work done by
J.~M.~Cline {\it et al.}, in which they found solutions with
$\theta_0\not=0$ ($\abs{\theta_0}\ll 1$)\cite{Cline}.
They incorporated the case of the explicit CP breaking, besides the 
spontaneous case, and their equation for $\theta$ corresponds to
that with $c=e=0$ in our analysis.\par
\vspace{36pt}
%\begin{center}
%{\bf Acknowledgment}
%\end{center}
%\vspace{24pt}
This work was partially supported by Grant-in-Aid for Encouragement of
Young Scientist of the Ministry of Education, Science and Culture,
No.07740224 (K.F.) and by Grant-in-Aid for Scientific Research Fellowship,
No.5106 (K.T.).
\vspace{40pt}
{\large\bf Figure Captions}\par\noindent
\begin{itemize}
\item[Fig.1:] The numerical solution of $\theta(y)$ for $\theta(0)=0.002$
 and $\theta(1)=0$. The parameters are $(b,c,e) = (3,7,7)$.
\item[Fig.2:] The numerical solution of $\theta(y)$ for $\theta(0)=1$
 and $\theta(1)=0$. The parameters are $(b,c,e) = (3,-1,0)$.
\item[Fig.3:] The numerical solution of $\theta(y)$ for 
 $\theta(0)=\theta(1)=0$. The parameters are $(b,c,d,e)=(3,12.2,-2,12.2)$.
\item[Fig.4:] The profile of the bubble wall corresponding to the solution
 in Fig.3 with $x = az$. 
 The solid line represents the kink, which is the absolute value
 of the VEV with the maximum being normalized to unity.
 The dashed line and dashed-dotted line are the real and 
 imaginary parts of it, respectively.
\item[Fig.5:] Contour plot of the chiral charge flux, normalized as
 $F_Q/(uT^3(Q_L-Q_R))$, for $u=0.1$ and $T=100$GeV, where $u$ is the 
 wall velocity and $F_Q$ is defined in \cite{FKOTTb}.
\item[Fig.6:] Contour plot of $V_{eff}/(a^2 v^2)$ along the kink, as a function
 of $y$ and $\theta$, for the parameters used to find the solution in
 Fig.3 and for $\beta=\pi/4$.
\end{itemize}
%
%
%     Figures
%
\epsfysize=\textheight
\centerline{\epsfbox{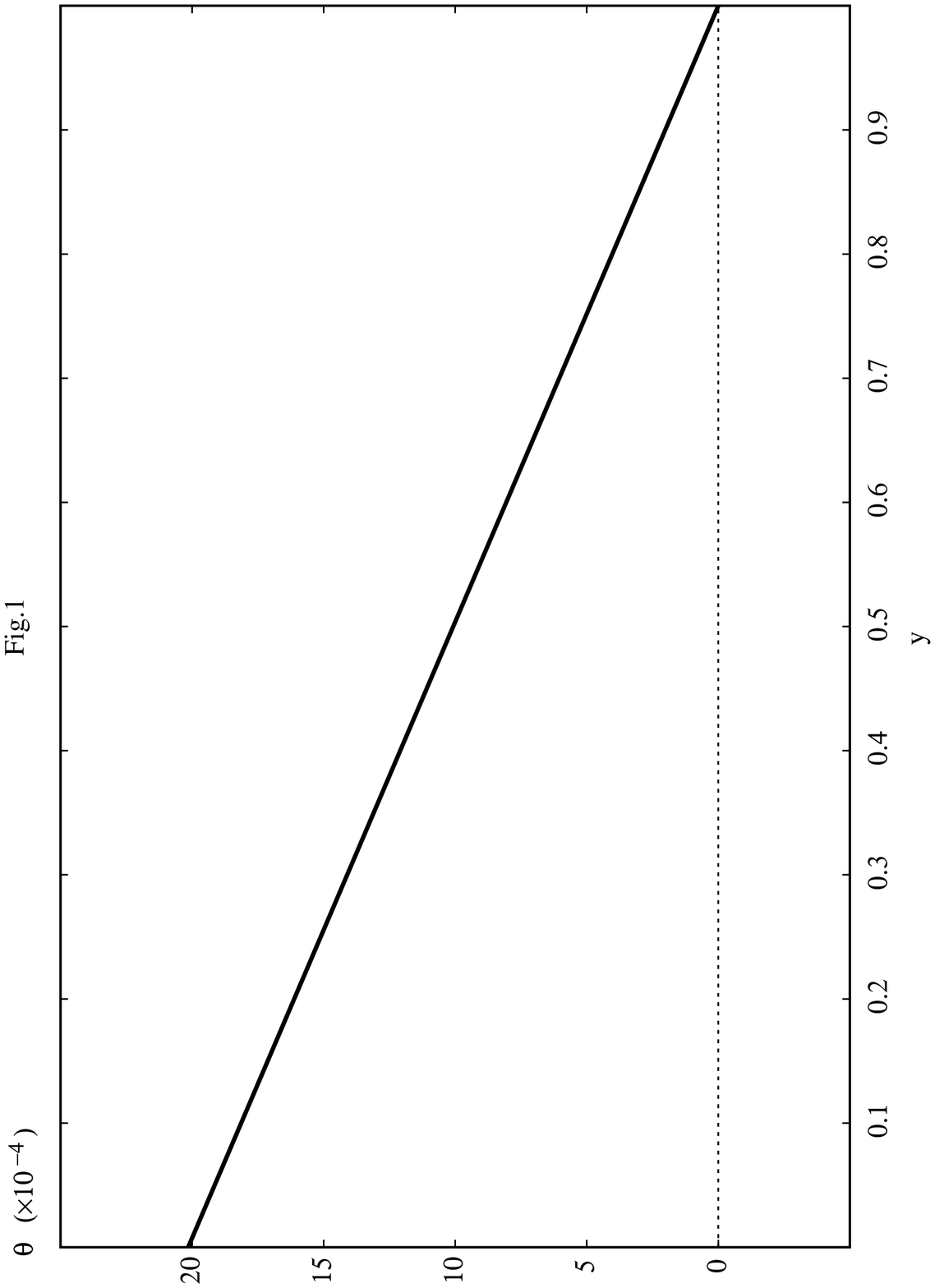}}
\newpage
\epsfysize=\textheight
\centerline{\epsfbox{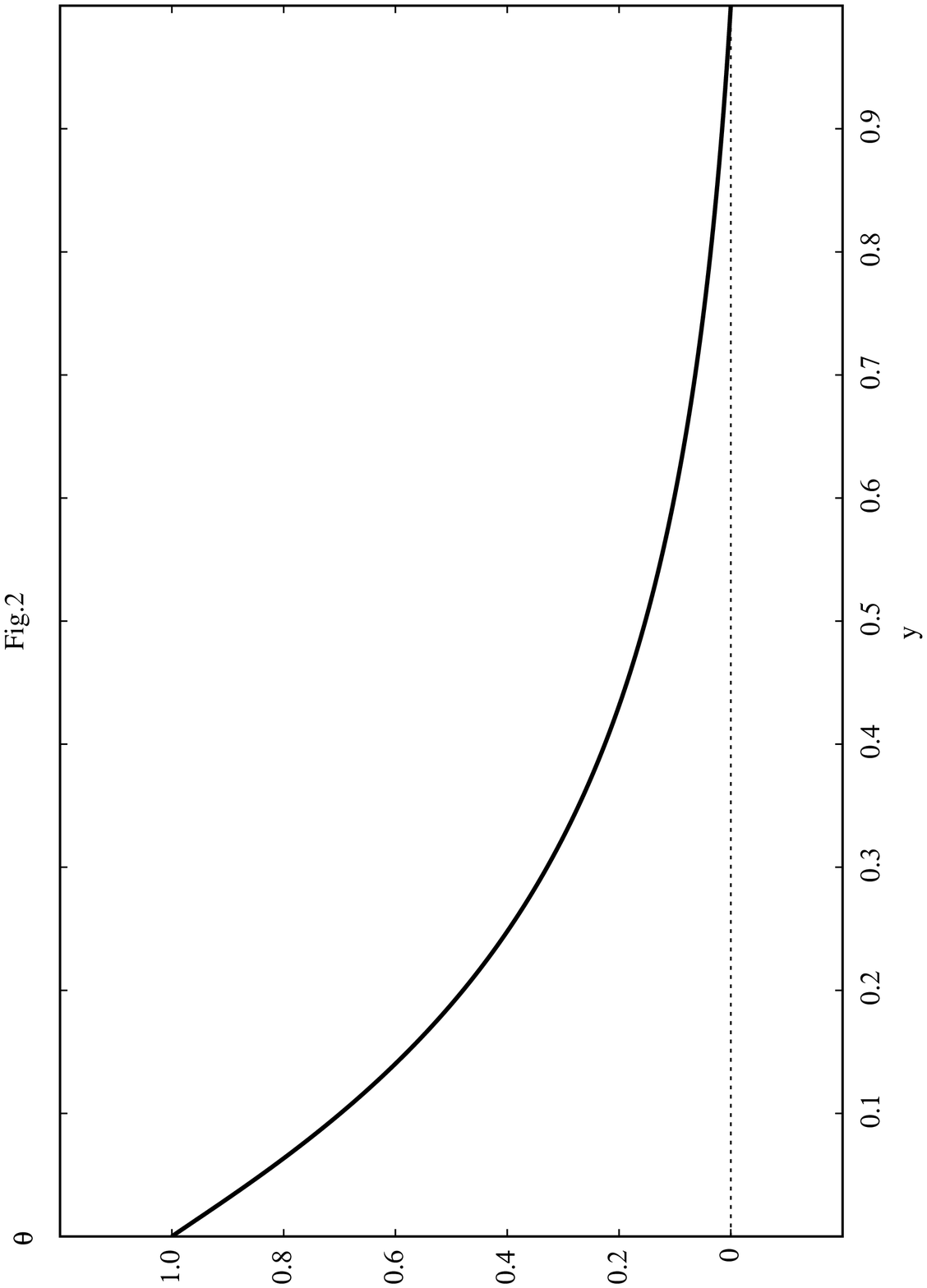}}
\newpage
\epsfysize=\textheight
\centerline{\epsfbox{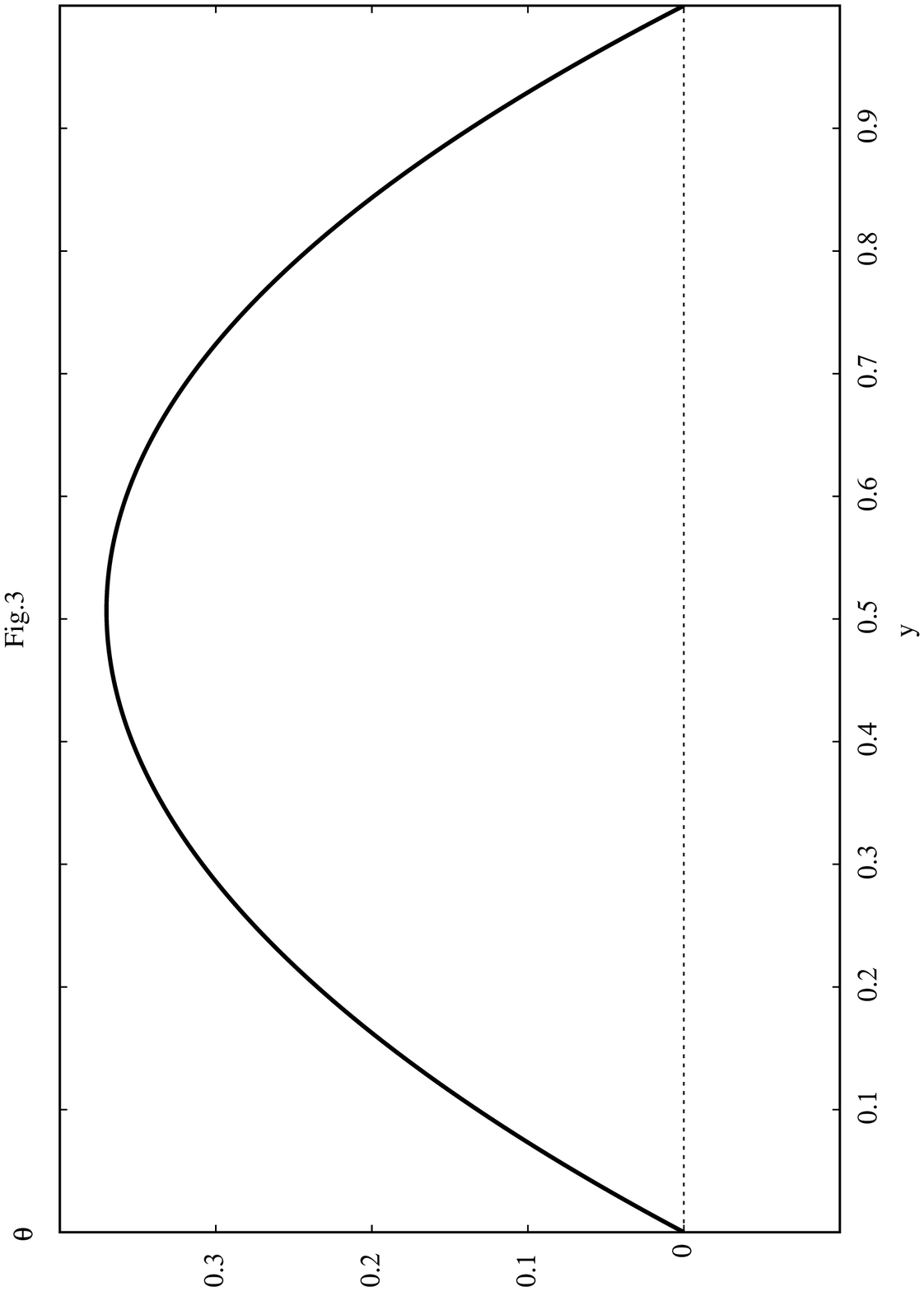}}
\newpage
\epsfysize=\textheight
\centerline{\epsfbox{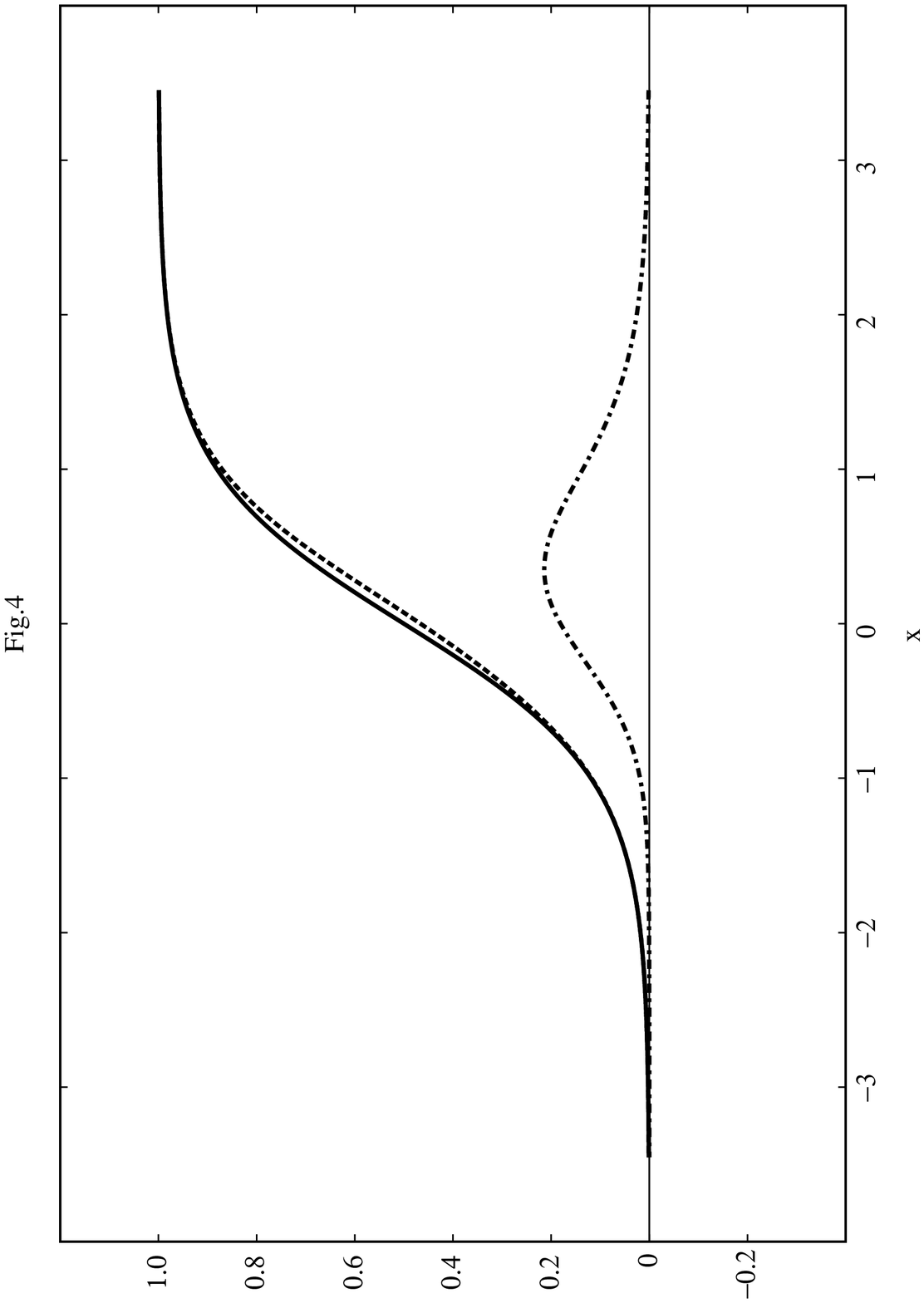}}
\newpage
\epsfysize=\textheight
\centerline{\epsfbox{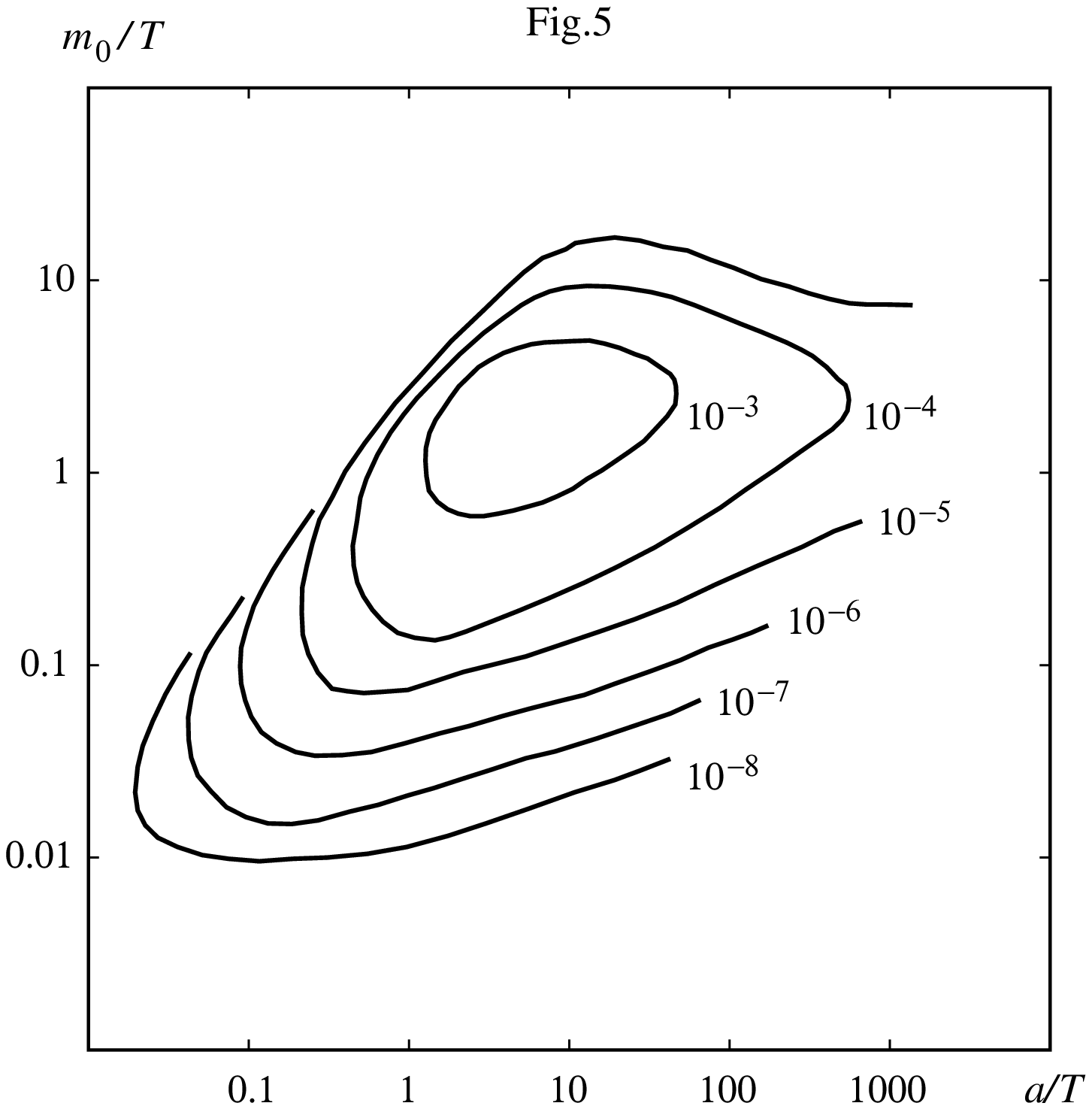}}
\newpage
\epsfxsize=\textwidth
\centerline{\epsfbox{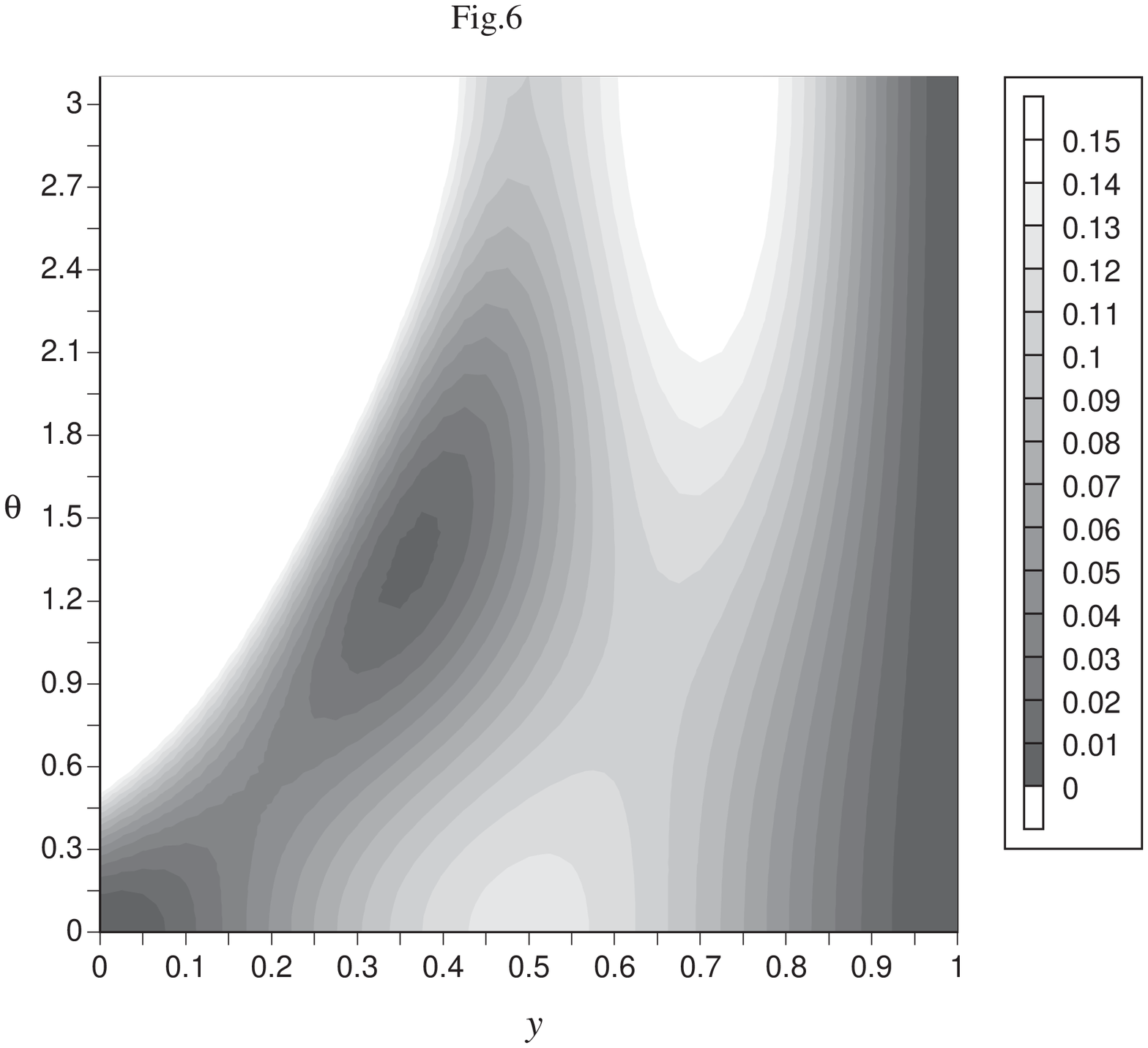}}
\end{document}